# Increase in the Absorption Length of Inclined Extensive Air Showers with Different Selection Types

E.S. Nikiforova

*Yu.G. Shafer Institute of Cosmophysical Research and Aeronomy,*
*31 Lenin Ave., 677980 Yakutsk,*

Presenter: E.S. Nikiforova (nikiforova@ikfia.ysn.ru), rus-nikiforova-E-abs1-HE14-oral

The dependence of the number of extensive air showers (EAS) on the zenith angle $\vartheta$ is obtained in groups with a constant solid-angle step with different selection types at separations between stations of 500 and 1000 m by using the Yakutsk EAS array data. A bend point is formed at the zenith angle ~50°. This point shifts with the increasing energy in the direction corresponding to the shift of the cascade-curve maximum depth with the variation of energy. The boundaries of change of these dependences obtained separately at the distances between stations of 500 and 1000 m are different that may be caused by the uncertainty of the lateral distribution function (LDF) used. Thereby, energies of showers detected by the stations with a different separation are not joined.

In [1] the presence of bent point depending on the number of extensive air showers (EAS) with $E_0>10^{18}$ eV on a zenith angle $\vartheta$ is shown. At first, as a zenith angle $\vartheta$ increases, the number of showers decreases or remains constant and then after reaching ~50° a considerable rise of them begins. The relation between a bend point and a boundary of change of charged particle absorption lengths at a distance of 600 m is also shown. With a rise of zenith angle the transition from the predominance of electron-photon shower component with a short absorption length to the mainly muon component with a weak absorption takes place.

At the Yakutsk EAS array the ground-based stations are located in the form of grid consisting of equilateral triangles with sides of 500 m – a small "master" and 1000 m – a large "master". Showers are registered by the array when signals from three stations forming a master triangle coincide. In present paper the dependence for different selection types of showers, as well as for showers registered separately by large and small "masters" are analysed.

Showers have been divided into groups with a constant step in $\cos\vartheta$ which corresponds to a constant solid angle [1]. The charged particle absorption length at a distance of 600 m from the axis $\lambda_{600}$ and the energy have been calculated by formulae accepted at the Yakutsk array [2,3]. Showers with axes located within the array perimeter have been selected.

Fig.1 presents two types of dependences of the number of showers in groups on a zenith angle $\vartheta$ at different energies for a small "master" by the following selection criterions: 1) with densities > 0.8 particles/m$^2$ and 2) with densities > 2.0 particles/m$^2$ at each of three stations forming a "master" triangle. A step for $\cos\vartheta$ has been taken as 0.025. On the plot a bend point in dependences at ~50°, its shift with the energy is seen. There is a restriction in large zenith angles for treated showers which had been changed during different years. In this case a bend point doesn't depend on a density at "master" stations, by which the showers have been selected.

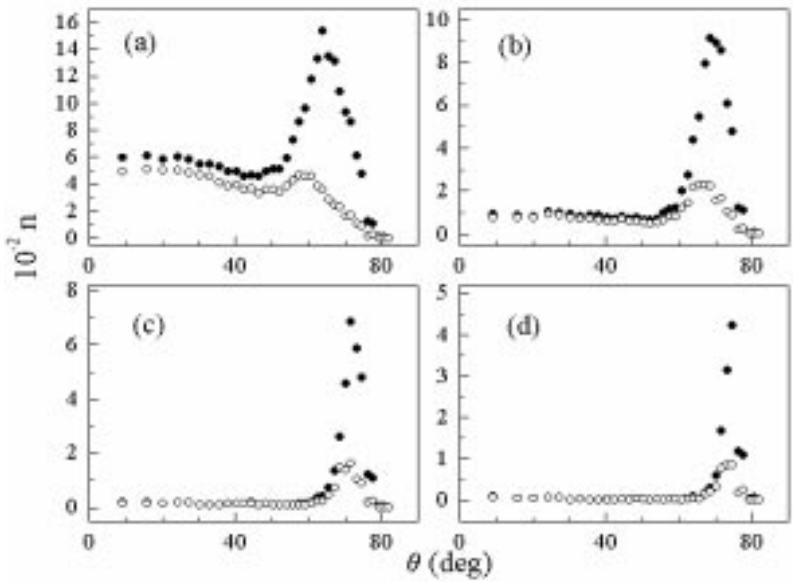

**Fig.1.** Dependence of n - the numbers of showers in groups on a zenith angle $\vartheta$ for a small "master" at following different energies: (a) - $\log(E_0) > 18.2$; (b) - $\log(E_0) > 18.6$; (c) - $\log(E_0) > 19.0$; (d) - $\log(E_0) > 19.4$ ($E_0$- in electronvolts). ● is a selection criterion of 0.8 particles/m², ○ is 2.0 particles/m².

The shower characteristics with zenith angles $\vartheta > 60°$ at the Yakutsk array have not studied. The energy determination formula extrapolation to such angles, apparently, leads to the systematic deviations. There are no reasons to consider that these deviations can lead to a bend in the dependence, moreover, on the given plots there is one bend shifting in dependence on the energy, and a sharp change of dependence at $\vartheta=60°$ is absent. For us in this case the general view is important, not absolute values of obtained dependences.

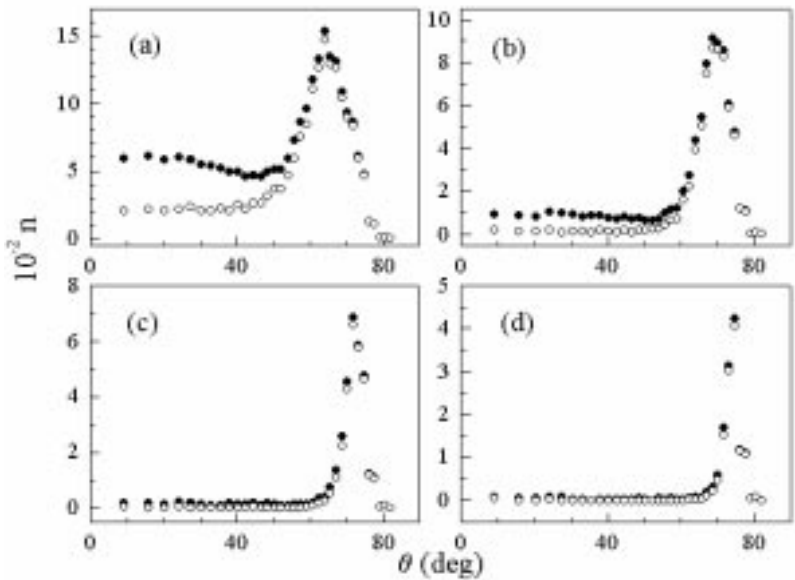

**Fig.2.** Dependence of n – the numbers of showers in groups on a zenith angle $\vartheta$ in selection criterion of 0.8 particles/m²; ● is for a small "master", ○ is only for a small "master" without a large one. Energies correspond to the given ones in Fig.1.

There are only the small number of showers, at registration of which a small and large "masters" simultaneously snap into action. For comparison in Fig.2 the dependences for showers are given in which: 1) a small "master" has snapped into action and 2) only a small "master" has snapped into action but a large one has not done. For the given plot the showers with densities greater then 0.8 particles/m² at each of three "master" stations have been selected. As is seen, in this case the dependence is not much affected by the account of a large "master".

Fig.3 shows the dependences analogous ones shown in Fig.1 for a large "master" beginning from higher energies in comparison with energies in Fig.1. A step for cosθ has been taken as 0.05. The bend of dependences for showers chosen in criterion of 2.0 particles/m² is less expressed in comparison with dependences in criterion of selection of 0.8 particles/m² because of a less statistics. In this case there is no any differences of the change boundary of absorption length for different types of selection.

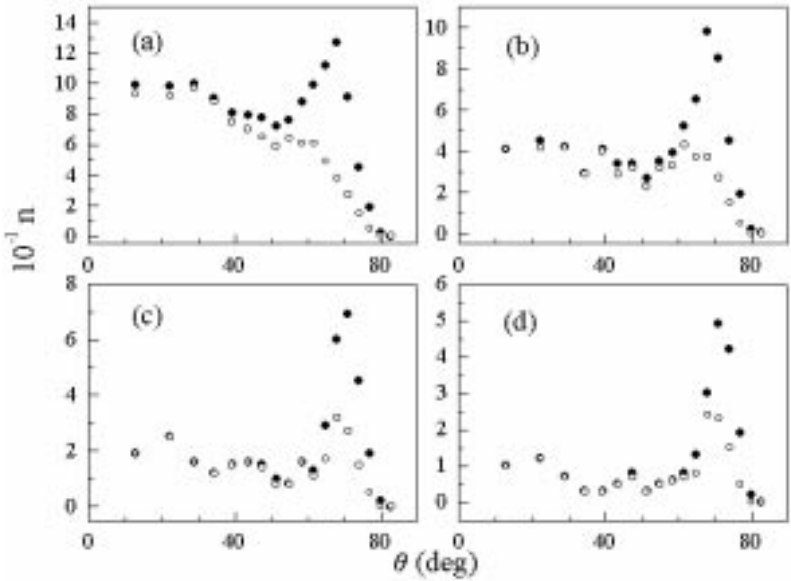

**Fig.3.** Dependence of n – the numbers of showers in groups on a zenith angle θ for a large "master" at following different energies: (a) - $\log(E_0) > 18.8$; (b) - $\log(E_0) > 19.0$; (c) - $\log(E_0) > 19.2$; (d) - $\log(E_0) > 19.4$ ● is a selection criterion of 0.8 particles/m², ○ is 2.0 perticles/m².

The selection of showers with densities greater than 2.0 particles/m² at each of three stations forming a "master" triangle is used at the Yakutsk array in studies of the energy spectrum and cosmic ray anisotropy. The selection of showers with densities greater than 0.8 particles/m² is also used in studies of anisotropy and results of analysis at such selection criteria are in agreement between each other. The showers selected in criterion of 0.8 particles/m² have been taken for the comparison of boundaries of the change of absorption length for large and small "masters" because of more statistics and absence of noticeable distinctions in comparison with a selection in criterion of 2.0 particles/m² in Fig.1 and 3.

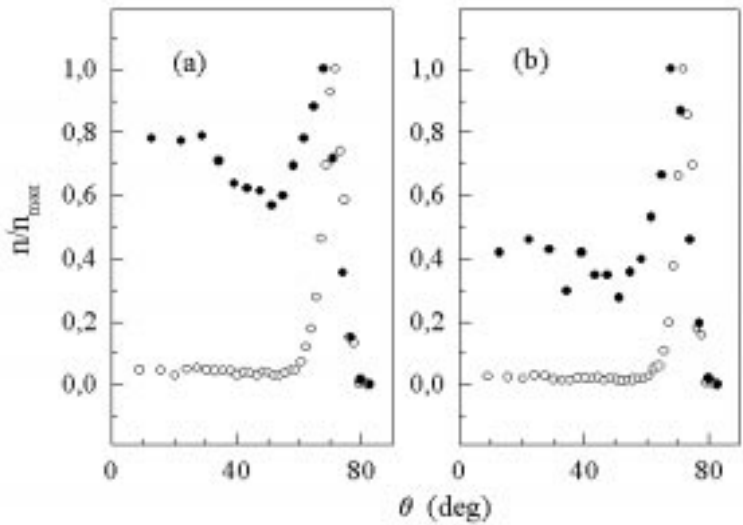

**Fig.4.** Dependence of $n/n_{max}$ – the numbers of showers in groups normalized by maximum on a zenith angle θ for the following energies: a) - $\log(E_0) > 18.8$; (b) - $\log(E_0) > 19.0$. A selection criterion is 0.8 particles/m² ● is a large "master", ○ is a small "master".

Distributions have been normalized by maximum. It is seen in Fig.4 that the bend points in distributions correspond to the essentially differing values of a zenith angle θ. The energies of a large "master" correspond to the less energies of a small "master". In this case the function of lateral distribution (LDF) will be flatter. One can suppose a somewhat different variant of explanation of the obtained contradiction. For showers close to the vertical the energies and LDF have been determined correctly, non-coinciding energies and LDF for the inclined showers require a review.

The work was financially supported by Ministry of Education and Science of Russia.